\documentclass[dvips]{article}

\usepackage{icrc2011}

\title{Observations of SNR CTA 1 and the Cyg OB1 region with VERITAS}

\newcommand{\etal}{\MakeLowercase{\textit{et al. }}} 
\shorttitle{Aliu \etal Observations of SNR CTA 1 and the Cyg OB1 region with VERITAS}

\authors{Ester Aliu$^{1}$ FOR  THE VERITAS COLLABORATION$^{2}$}
\afiliations{$^1$Barnard College, 3009 Broadway, 10027, New York City, US\\ 
$^2$ see Holder et al (these proceedings) for a complete list of authors }
\email{ealiu@astro.columbia.edu}

\abstract{The Cygnus region is a nearby very active star forming region, containing several OB associations, considered as tracers of young pulsars. Above 12 TeV, the Milagro Collaboration has reported the discovery of a very large source, MGRO J2019+37, lying towards the Cyg OB1 association, at the edge of the Cygnus region. The young and energetic pulsar PSR J2021+3651 has been proposed to power this emission. We present here the result of deep VERITAS observations of this region at energies above 650 GeV. These observations unveil extended and complex TeV emission compatible with MGRO J2019+37, likely made of multiple sources, and a clearly separated point source emission from the direction of CTB 87, a pulsar wind nebula candidate.  We will also report on the detection of TeV emission from the young Galactic SNR CTA 1, likely powered by the first pulsar discovered through its gamma-ray radiation.}


\begin{document}
\maketitle

\section{Introduction}
One of the key science goals of the study of the very high energy (VHE, $>$100 GeV) sky is to locate the sources responsible for the bulk of the Galactic cosmic rays.~Those can generate gamma rays via the decay of neutral pions produced in hadronic interactions with the ambient material.~Nevertheless, relativistic electrons can also produce gamma rays by inverse Compton (IC) scattering on ambient microwave, IR, or optical photons.~The measured properties of the VHE sources such as spectrum and morphology can help to disentangle between the leptonic and hadronic process along with the study of the lower energy counterparts. 
\par
Among the Galactic TeV sources found thus far there are supernova remnants (SNR), pulsar wind nebulae (PWN), binary systems and, also, unidentified objects (UNID), found in the surveys of the Galaxy\footnote{See TeVCat, an online catalog for TeV Astronomy, http://tevcat.uchicago.edu}.~About a 10\% efficiency of cosmic ray acceleration in SNRs, assuming the typical SN explosion kinetic energy release of 10$^{51}$ erg, would compensate the cosmic ray losses in the Galaxy~\cite{ginzburg64}.~This is the reason why SNRs have been considered the major source of Galactic cosmic rays, although this lacks a definitive proof.~In this paper, we report the detection by VERITAS of an extended VHE emission from the young Galactic SNR CTA 1 and unveil a complex VHE emission region towards the Cygnus region, one of the nearest and more active star forming regions in the Galaxy.

\section{The VERITAS array}
VERITAS is an array of four imaging atmospheric Cherenkov telescopes (IACT) located at the Fred Lawrence Whipple Observatory in southern Arizona, USA~\cite{holder06, holder11}.~Each telescope has a mirror area of 110 m$^2$ and is equipped with a 499-pixel camera of 3.5$^{\circ}$ diameter field-of-view (FoV).~The system, completed in the fall of 2007, is run in a coincident mode requiring at least two of the four telescopes to trigger in each event.~This design enables the observations of astrophysical objects in the energy range from 100 GeV to 30 TeV.~VERITAS has a good energy (15-20\%) and angular ($<$ 0.1$^{\circ}$) resolution,  and a 5$\sigma$ point-source sensitivity of 1\% of the Crab Nebula flux above 300 GeV in less than 26 hr observation at a 20$^{\circ}$ zenith angle. 
\par
For the analysis presented here, events were reconstructed using the classical Hillas parameters~\cite{hillas85} for the case of stereoscopic observations (see e.g.~\cite{krawczynski06}).~The $\gamma$-ray event selection was optimized for a hard-spectrum source at 1\% of the Crab Nebula flux.~The search for the $\gamma$-ray signal is performed with two fixed integration radii ($\theta_{int}$ = 0.09$^{\circ}$ and $\theta_{int}$ = 0.23$^{\circ}$), to account for the possibility of finding both point-source and extended emission.  

\section{The SNR CTA 1}
CTA 1 is a composite SNR with a shell-type structure in the radio band with a diameter of 1.8$^{\circ}$ and a center filled morphology in X-rays corresponding to a PWN~\cite{slane97}.~The radio shell is incomplete towards the NW of the remnant, possibly to rapid expansion of the shock into a lower density region in the NW,~as supported by HI observations~\cite{pineault93}.~The system has a derived distance of $\sim$1.4$\pm$0.3 kpc and an estimated age of $\sim$1.3$\times$10$^{4}$ yr.~The discovery of an X-ray jet in the brightest part of the central X-ray emission provided unambiguous evidence of the presence of an energetic pulsar~\cite{halpern04}.~Pulsations have only been recently found through the HE ($>$100 GeV) gamma-ray band using the LAT data~\cite{abdo08}, making this the first pulsar discovered by its gamma-ray radiation.~The same authors determine the spin-down power to be $\sim$4.5$\times$10$^{35}$ erg s$^{-1}$. 
\par
The presence of both a bright PWN and a SNR shell interacting with the ambient ISM,  makes CTA 1 a prime candidate for VHE gamma-ray observations. VHE emission from the nebula has been predicted in one study to be 10\% of the Crab Nebula flux, which should be detectable by VERITAS~\cite{zhang09}.

\subsection{Observations, results \& discussion}
The observations of CTA 1 with VERITAS span from October 2010 to January 2011, and amount to 26.5 hr, after selection for good weather and absence of hardware problems.~Wobble mode observations around the nominal position of CTA 1 were used with an offset of 0.7$^{\circ}$, motivated by the large size of the remnant.~An extended VHE source centered around the pulsar and embedded in the large SNR is found as shown in Fig.~1.~The detected signal has a pre-trials statistical significance of 7.3 standard deviations ($\sigma$), and 6.2$\sigma$ post-trials.\\
 \begin{figure}[h]
  \vspace{5mm}
  \centering
  \includegraphics[width=3.in]{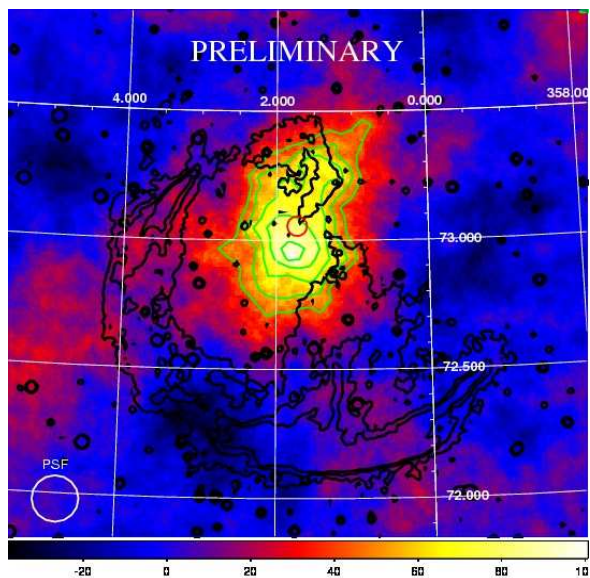}
  \caption{Excess-counts sky map from the SNR CTA 1 as seen by VERITAS. The radio emission from the shell is indicated with the black contours. VERITAS 3 to 7$\sigma$ significance contours are shown in green \label{fig:CTA1}}
 \end{figure}
\par
The observed morphology of the VHE emission suggests a young PWN. A more complete presentation of the results, including spectra and flux, and its interpretation will be shown at the conference.

\section{ The Cyg OB1 association }
The Cygnus region is known to be one of the nearest ($\sim$ 1.7 kpc) and more active regions of creation and destruction of massive stars in the Galaxy, containing numerous HII regions, Wolf-Rayet stars, OB associations.~It is therefore a good location to study the origin of the Galactic cosmic rays through gamma-ray observations.~First surveys of the HE gamma-ray sky with EGRET found $\leq$10 UNID sources in the Cygnus region~\cite{hartman99}.~A statistically significant correlation of these sources near the Galactic plane with OB associations, considered as tracers of young pulsars, was found~\cite{kaaret96}.
\par
Most recently, the Milagro water cherenkov detector has performed a large-scale survey of the Northern sky at much higher energies ($>$12 TeV) discovering a population of extended sources in the Cygnus region without compelling counterparts~\cite{abdo07}.~The brightest and largest source reported by the Milagro Collaboration was MGRO J2019+37, overlapping with the Cyg OB1 association, a less massive OB association than the nearby Cyg OB2.~MGRO J2019+37 is located with an accuracy of $\pm$0.4$^{\circ}$. The best-fit ellipse of MGRO J2019+37 covers an area of 0.6$^{\circ}$ $\times$ 1.0$^{\circ}$, although the observed emission extends beyond the ellipse.~The total measured flux is about 80\% of the Crab Nebula flux above 12 TeV.
\par
The origin of the gamma-ray emission from MGRO J2019+37 has been subject of much speculation.~Emission from the shocks driven by the WR stars contained in the young cluster Ber 87 in the Cyg OB1 has been proposed~\cite{bednarek07}.~Radio surveys to find other plausible counterparts have also been performed~\cite{paredes09}.~However, the PWN produced by the energetic young pulsar PSR J2021+3651 ($\dot{E}$ = 3.4$\times$ 10$^{36}$ erg/s) is still a common interpretation, given the observed offset between the emission maximum and the pulsar, an effect often seen in TeV PWNe. 

\subsection{Observations \& results}
VERITAS observations towards the Cyg OB1 region took place between May 2010 and December 2010, and amount to 75 hours after selection for good weather and absence of hardware problems.~Wobble mode observations around PSR J2021+3651 were used in the first half of the observations with an offset of 0.7$^{\circ}$.~Motivated by the initial results, the wobble position was subsequently moved $\sim$0.5$^{\circ}$ to the location of the hard X-ray integral source IGR J20188+3647.
\par
Two source regions were defined {\it a priori} and therefore excluded from the background estimation. One source region was defined around the location of CTB 87, located only 1.1$^{\circ}$ from PSR J2021+3651, and a good candidate for TeV emission. Also, an extended source region corresponding to the best-fit ellipse of MGRO J2019+37 was defined.~Bright stars in the FoV were also excluded from the background estimation. 
\par
Results are shown in Fig 2 and Fig 3, for the point-source search and extended emission search, respectively, as defined in Sec~2.~The point-source search reveals significant (6.1$\sigma$ post-trials) emission at the location  of CTB 87.~The centroid of the TeV emission is used to assign the name VER J2016+372 to it. In the larger search window a significant (7.4$\sigma$ post-trials) elongated and extended emission region, compatible with the best-fit extension of MGRO J2019+37, is found.~This extended emission appears as a complex region of TeV emission in the point-source search, with pre-trial statistical significances in the different hotspots from 3$\sigma$ to 6.2$\sigma$.
\begin{figure}[t]
  \vspace{5mm}
  \centering
  \includegraphics[width=3.25in]{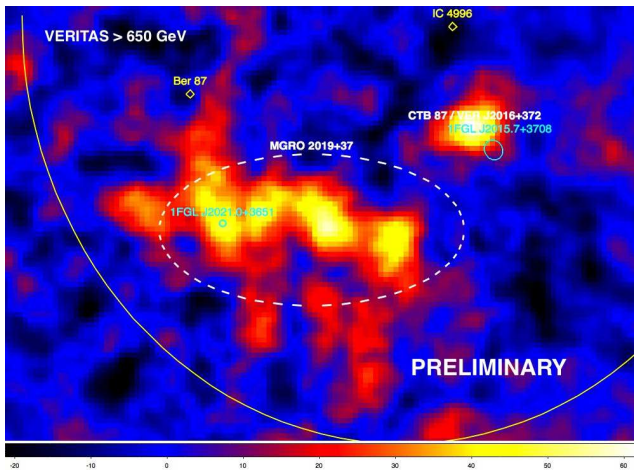}
  \caption{VERITAS excess-counts map towards Cyg OB1 region (big yellow circle) above 650 GeV. This skymap has been produced with to the point source search window ($\theta_{int}$=0.09$^{\circ}$, see text). The positions of Ber 87 and IC 4996 (young clusters associated with Cyg OB1), are indicated, along with those of 1FGL sources in the region. The best-fit ellipse of MGRO J2019+37 is indicated with a white dashed line. \label{fig:pointsourcesearch}}
  \label{simp_fig}
 \end{figure}
  \begin{figure}[t]
  \vspace{5mm}
  \centering
  \includegraphics[width=3.25in]{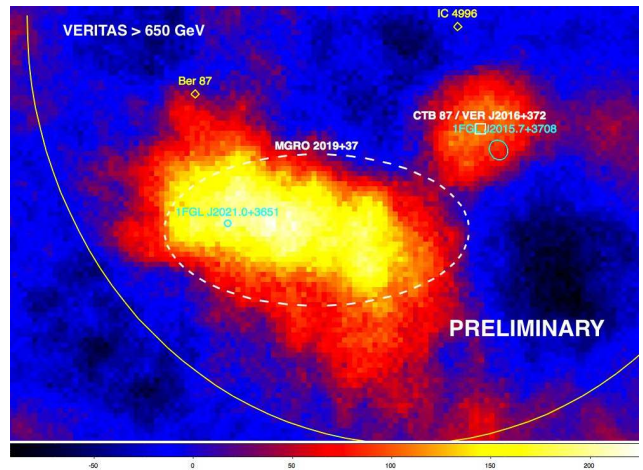}
  \caption{VERITAS excess-counts map $>$ 650 GeV towards Cyg OB1 region using the extended emission search window ($\theta_{int}$=0.23$^{\circ}$).~Look at Fig.~1 caption for further information. \label{fig:extendedsourcesearch}}
  \label{simp_fig}
 \end{figure}
\par
A preliminary flux of 1\% of the Crab Nebula flux above 1 TeV and a hard spectrum with photon index $\Gamma\sim$2.1$\pm$0.4$_{stat}$ has been estimated for VER J2016+372/CTB 87.~The significance of this new VHE source increases with time as expected for a steady source.~A complete presentation of the results, including spectra and fluxes of all significant excess will be presented at the conference. 

\subsection{Discussion}
Along the line-of-sight of Cyg OB1 there are various spiral arms of the Galaxy, and therefore many young stellar objects associated with different arms could be concentrated on the sky.~This is for example the case of CTB 87, with a measured distance of d$\sim$ 6 kpc is located in the Perseus arm.~CTB 87 is known to be powered by a young pulsar, although the properties and energetics of this pulsar are not yet known.~The location of the compact VHE source VER J2016+372 is compatible with the position of CTB 87, and we can exclude VHE emission from the blazar with 99\% probability, as shown in Fig 4.~Also, the measured spectrum and the absence of variability throughout the observations are properties similar to other PWNe previously detected in the VHE band.~The LAT source 1FGL J2015.7+3708 is compatible with the position of CTB 87, but its high degree of variability favors an association with the blazar BS 2013+370~\cite{mukherjee00}. 
\par
On the other hand, the bright central VHE emission contains PSR J2021+3651 that has been for many years the only young and energetic pulsar known in this part of the Cygnus region~\cite{roberts02}.~It is also the brighter HE gamma-ray emitter in the FoV~\cite{halpern08, abdo09a}, and has been considered to be the powering source of  MGRO J2019+37.~These new VERITAS observations show that MGRO J2019+37 is a complex TeV-emitting region, likely powered by multiple sources.~As indicated in Fig 2, among the possible candidates,  because of positional coincidence, are the PWN of PSR J2021+3651, the Wolf-Rayet star WR 141, the transient IGR J20188+3647, the HII region Sh-104 and probably new cosmic rays sources not yet known.~An example of the latter is the discovered extended and non-thermal source in archival XMM data to the North of the HII region~\cite{zabalza10}, which clearly overlaps with the VERITAS emission as shown in Fig 4.~Deeper VERITAS observations are underway to try to disentangle the possible several sources in the Cyg OB1 and have a more accurate position for them.

\begin{figure}[t]
  \vspace{5mm}
  \centering
  \includegraphics[width=3.0in]{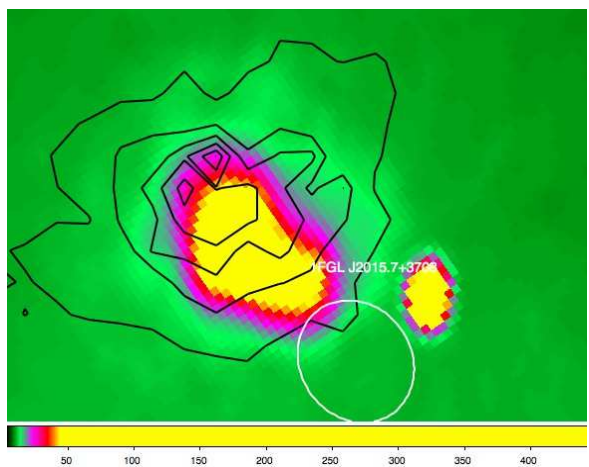}
  \caption{Radio image of CTB 87 with CGPS (1420 MHz).~Another bright radio source appears at only 11 arcmin from the PWN, which correspond to the blazar BS 2013+370.~Black contours indicate the observed VHE gamma-ray significance from 3 to 7$\sigma$.~1FGL J2015.7+3708 is a highly variable HE source (V=139~\cite{abdo10}) \label{fig:CTB87radio}}
  \label{simp_fig}
 \end{figure}
  \begin{figure}[t]
  \vspace{5mm}
  \centering
  \includegraphics[width=3.0in]{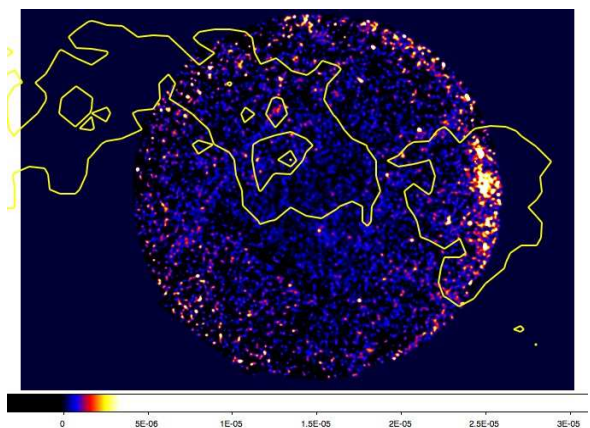}
  \caption{In this archival XMM observation dedicated to search for counterparts for the hard X-ray transient IGR J20188+3647, an extended and non-thermal X-ray source is found on the edge of the FoV~\cite{zabalza10}. This UNID source overlaps with the VERITAS emission, indicated with the yellow contours (from the point source search).\label{fig:XMM_VERITAS}}
  \label{simp_fig}
 \end{figure}

\section{Conclusion}
VERITAS has detected extended VHE gamma-ray emission from  SNR CTA 1 and had resolved the mysterious multi-TeV source MGRO J2019+37 into a point-source emission from SNR CTB 87 and a complex region in the central and brightest part of MGRO J2019+37, the Cyg OB1 TeV complex. The origin of the TeV emission in both, CTA 1 and CTB 87 is likely nebular emission from a central and energetic pulsar, and therefore a leptonic process for the gamma-ray production is the most favorable.~Various sources could contribute in the extended emission seen towards the Cyg OB1 region.~Hints of that are present in the X-ray and HE gamma-ray data.~However, the main cosmic accelerators and, therefore the gamma-ray acceleration mechanisms of this complex VHE gamma-ray emitting region, remain unidentified.~Multiwavelength observations and more VHE gamma-ray data with VERITAS will be necessary to solve these new mysteries. \\

{\bf Acknowledgement} \\
This research is supported by grants from the US Department of Energy, the US National Science Foundation, and the Smithsonian Institution, by NSERC in Canada, by Science Foundation Ireland, and by STFC in the UK. We acknowledge the excellent work of the technical support staff at the FLWO and the collaborating institutions in the construction and operation of the instrument. E.A. acknowledges support through the Beatriu de Pinos Program of the Generalitat de Catalunya and  the partial support by the NSF grant Phy-0855627 at Barnard College.

\end{document}